Title: Photoinduced electrification of solids. IV. Space-charge effects assessed
Author: M. Georgiev (Institute of Solid State Physics, Bulgarian Academy of Sciences,
  1784 Sofia, Bulgaria)
Comments: 6 pages including 2 figures, all pdf format
Subj-class: cond-mat

Recent preprints have described the experimental evidence for the universal occurrence of short circuit photocurrents on illumination of solid state surfaces by strongly absorbed light. A likely mechanism has been proposed based on the photodesorption of surface ions to the surrounding atmosphere. Analyses have been made of observed oscilloscope tracings of short circuit photo voltages in terms of the linear response theory in fair concomitance. Space charge effects having been left unaccounted for, we now work out an approach to them.

1. Introduction

In a series of three arXiv preprints, we told of an unconventional photo charging effect which occurred virtually universally on solid state surfaces from insulators through metals through high-$T_c$ superconductors through natural biosamples [1-3]. Essentially photo voltages within the $\mu$V range, as induced by chopped laser light, were measured under short-circuit conditions [1]. Oscilloscope tracings at room temperature (RT) revealed a photo charging process with a time constant of the order of 1 $\mu$s under light on followed by a slower signal decay in the dark [2]. Such experiments were carried out on solid substances within the temperature range from liquid nitrogen temperature (LNT) to RT [3]. They all exhibited a thermally stimulated peak (TSP) near 200 K thermally activated at ~ 50 meV. This estimate as well as its universal character suggested that the peak possibly originated from emptying of surface traps. An analysis of the oscilloscope tracings in terms of linear rate equations indicated a close relationship [2]. Based on the collected data, a model was confirmed where photo desorption played the major charging role by inciting ion detachment from the sample surface into the surrounding atmosphere.

Nevertheless the simple linear theory, in spite of being justified by its application to surface rather than bulk reactions, was short of disclosing the role of subsurface space-charge effects in shaping the photo voltage response of the sample. Methods for dealing with space charges have been thoroughly reviewed by Sigmond [4] for the unipolar corona charge case. Although he has also not included an irradiation source (illumination source in our case, to be exact), we consider some of his analyzes presently applicable too. In accordance, we will try to account for the subsurface space charges and see what else can be obtained from the experimental data on that basis.

2. Basic equations

Introducing the electrostatic potential $\phi(\mathbf{r})$ and the field strength $\mathbf{E}(\mathbf{r}) = -\nabla\phi(\mathbf{r})$, the unipolar space-charge equations read:

$$\nabla \bullet \mathbf{E} \equiv -\Delta\phi = \rho/\varepsilon_0 \text{ (MKS)} = 4\pi\rho/\varepsilon_0 \text{ (CGSE)} \qquad \text{(Poisson's equation)} \qquad (1a)$$

$$\mathbf{j} = \mu\rho\mathbf{E} - D\nabla\rho \qquad \text{(ion current density)} \qquad (1b)$$

$$D = \mu k_B T/e \equiv U_D \mu \qquad \text{(Nernst-Einstein-Townsend equation)} \qquad (1c)$$

$$\nabla \bullet \mathbf{j} + \partial\rho/\partial t = 0 \qquad \text{(continuity equation)} \qquad (1d)$$

in which $\rho$ is the space charge, $\mu$ is the mobility, D is the diffusion coefficient, $\varepsilon_0$ is the dielectric constant. If we discard the ionization (source) terms ($\partial\rho/\partial t = 0$) the system (1) reduces to a single equation of the form

$$\nabla\phi \bullet \nabla(\Delta\phi) + (\Delta\phi)^2 - U_D \Delta(\Delta\phi) = 0 \qquad (2)$$

while on neglecting diffusion as well we get

$$\nabla\phi \bullet \nabla(\Delta\phi) + (\Delta\phi)^2 = 0 \qquad (3)$$

It will be seen that apart from ionization the basic difficulties on solving analytically the space charge problem (2) arise from the diffusion term $U_D \Delta(\Delta\phi)$.

An extension of the unipolar equations into bipolar ones can be made straightforwardly by assuming $\mu_+ \approx \mu_- = \mu$ as we did while discussing the surface photo charge effect [2]. Using $\Delta\phi = -\rho/\varepsilon_0$ we obtain from equation (3): $\nabla\phi \bullet \nabla\rho = \rho^2/\varepsilon_0$ and hence equation (3) turns in $\nabla\phi \bullet \nabla(\varepsilon_0/\rho) \equiv -\mathbf{E} \bullet \nabla(\varepsilon_0/\rho) = -1$. For a specific sample geometry, we get $\varepsilon_0\{E_Z[1/\rho(Z)] - E_{Z0}[1/\rho(Z_0)]\} = (Z - Z_0)$ where the zero-labeled values relate to the surface, otherwise to the bulk. Multiplying by the mobility $\mu$ the last equation transforms into $\varepsilon_0\{\mu E_Z[1/\sigma(Z)] - \mu E_{Z0}[1/\sigma(Z_0)]\} = (Z - Z_0)$ where $\sigma$ is the dc conductivity. For an uniform field E along Z we also get

$$1/\rho(Z) - 1/\rho(Z_0) = \mu t / \varepsilon_0 \qquad (4)$$

where $t = \Delta Z / |\mu E|$ is the drift time. Equation (4) has been obtained and discussed elsewhere [4]. Being derived under fairly general assumptions though discarding diffusion and ionization, yet it is of some use for the present purpose. Nevertheless it is significant for showing just how the theory works. In particular, the solution emerging from equation (4) is

$$\rho = \rho_0 / [1 + (t/\tau_d)] \qquad (5)$$

where

$$\tau_d = \varepsilon_0 / \mu\rho_0 \text{ (MKS)} = 4\pi\varepsilon_0 / \sigma_0 \text{ (CGSE)} \qquad (6)$$

is the dielectric (Maxwell's) relaxation time. For an estimate, we take $\rho_0 = \frac{1}{2} \times 10^{15}$ cm$^{-3}$ (half filling of the surface sites) $\times 4.8 \times 10^{-10}$ CGSE = $2.4 \times 10^5$ CGSE cm$^{-3}$. Inserting tentatively $\mu = 1$ cm$^2$/Vs = 300 CGSE and $\varepsilon_0 = 4$, we get $\tau_d \sim 4$ ns. Actually the dielectric relaxation time is somewhere between 1 ns $\div$ 1 ms depending on the excitation light intensity. Indeed we set $\rho_0 = \eta k I_0 \tau$ where $\eta$ is the ionization yield, k the absorption constant, $I_0$ the excitation light intensity, $\tau$ an appropriate time-response constant.

Setting $Z = t\mu|E_Z|$ in (5) and inserting into Poisson's equation (1a) we get for a specific geometry $d^2\phi/dZ^2 = -\rho/\varepsilon$ and integrate it to $(d\phi/dZ) - (d\phi/dZ)_0 = -\int_0^Z \rho_0 [1 + t/\tau_d]^{-1} dZ = -\rho_0 \mu|E_Z|\tau_d \ln(1 + t/\tau_d)$ which is solved to $(d\phi/dZ) = (d\phi/dZ)_0 / [1+ \ln(1 + t/\tau_d)]$ and then to $\phi(Z) - \phi(Z_0) = -(\varepsilon_0 E_Z^2/\rho_0)[(1+ t/\tau_d) \ln(1 + t/\tau_d) - 1]$ where $E_Z > 0$. We obtain concomitantly

$$\phi(Z) - \phi(Z_0) = (\varepsilon_0 E_Z^2/\rho_0)[1 - (1+ t/\tau_d)\ln(1 + t/\tau_d)] \equiv \phi(Z_0)[1 - (1+ t/\tau_d)\ln(1 + t/\tau_d)] \quad (7)$$

and display in Figures 1(a) through 1(b) graphs of the space-charge and electrostatic potential drift-time axis distributions according to equations (5) and (7), respectively.

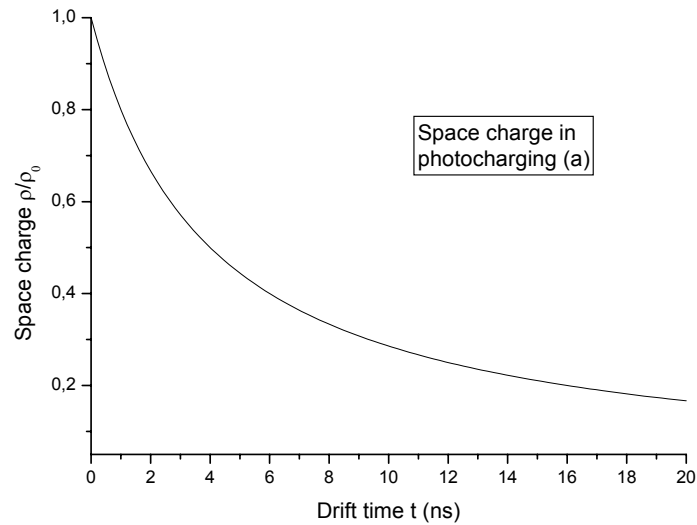

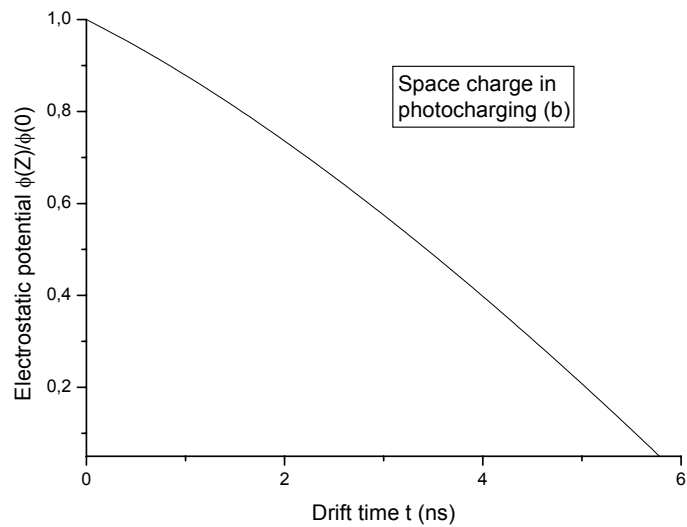

Figure 1. Drift-time axis distributions from equations (5) and (7), respectively, of the space charge (a) and the electrostatic potential (b). The drift time is $t = Z/\mu E_Z$, Z is the bulk depth below the illuminated surface at $Z_0 (= 0)$, $\mu$ is the mobility of carriers, $E_Z$ is the electric field

## 3. Link between surface and bulk processes

At this point we need calculating the balance between bulk and surface charges. From equation (5) the integrated bulk charge at Z is:

$$P(Z) = \int_0^Z \rho(Z)dZ = \rho_0 \int_0^Z [1+(Z/\mu|E_Z|\tau_d)]^{-1} dZ = \rho_0 \mu |E_Z| \tau_d \ln(1+ Z/\mu|E_Z|\tau_d) \qquad (8)$$

on the assumption that

$$E_Z = -(d\phi/dZ)_0 / [1+ \ln(1 + t/\tau_d)] \qquad (9)$$

is only slowly dependent on Z. $E_Z(\infty)$ is seen to be vanishing, as it should, far away from the illuminated surface at $Z_0 = 0$ (t = 0). At $Z = \infty$, $P(\infty) = -(d\phi/dZ)_0 \rho_0 \mu \tau_d = j_0 \tau_d = Q_0$, the charge carried across the surface into the bulk.

Throughout this paper we adopt the following experimental geometry: An excitation beam of strongly absorbed light falls onto the sample surface at $Z_0 = 0$. It generates (sub)surface electronic carriers in either direct or indirect photo excitation processes which make that surface charged against the bulk. Within a short time $\tau_d$ (Maxwell's relaxation) a photo equilibrium is established between surface and bulk which gives rise to steady state values of the characteristic quantities, such as the space charges, differential and integrated, the electrostatic potential, and the electric field strength. This photo equilibrium is active as long as the excitation persists and ends up in time when the photo excitation ceases, the system returning to its initial state prior to the excitation. If, however, the configurational structure of the sample is altered by the photo excitation by means of photo induced reactions the new equilibrium state which is attained when the excitation is ceased may not reproduce the initial equilibrium prior to the excitation and the new equilibrium may exhibit residual features.

Let each balance, initial and residual, be in thermal equilibrium with the host material. For a binary system of photo excited carriers assuming Boltzmann statistics, we postulate

$$\rho^\pm(Z) = \rho^0 \exp(\pm e\phi(Z)/k_B T) \qquad (10)$$

Inserting into Poisson's equation (1a) we obtain the Poisson-Boltzmann (PB) equation:

$$\nabla^2[e\phi(Z)/k_B T] \equiv d^2[e\phi(Z)/k_B T]/dZ^2 = \kappa^2 \sinh(e\phi(Z)/k_B T), \qquad (11)$$

also represented in the form [for $\kappa^2 = (8\pi\rho_0 e/\varepsilon_0 k_B T)$]

$$\nabla^2 \psi(\zeta) \equiv d^2\psi/d\zeta^2 = \sinh(\psi) \qquad (12)$$

where $\psi(Z) = e\phi(Z)/k_B T$ and $\zeta = \kappa Z$ are the normalized potential and in-depth coordinate, respectively. The latter nonlinear equation has aroused some skepticism over the years related to the difficulties of proving its statistical compatibility. Only its reduced form at $e\phi/k_B T \ll 1$, known as Debye's equation, in which $\sinh(e\phi/k_B T) \approx e\phi/k_B T$ has been found compatible so far. Nevertheless, the PB equation has been used for drawing essential conclusions regarding the equilibrium space charge layers by ionic defects in ionic crystals [5]. Both spatially damping and periodic 1D solutions by elliptic functions have been derived and their relevance to physical problems studied [6]. We stress that equations (5) and (10) are characteristic of the photo equilibrium and the thermal equilibrium of ionic charged carriers, respectively. One way or the other, the aperiodic solution to the PB equation (11) reads [5,7]:

$$\psi(\zeta) = \ln\{\coth[(\zeta/2) + \beta]\}^2 \qquad (13)$$

This solution describes the distribution of the space-charge layer beneath the crystalline surface. From equation (13) we have

$$\psi(\zeta)/\psi(0) = \ln\{\coth[(\zeta/2) + \beta]\}^2 / \ln\{\coth[\beta]\}^2 \qquad (14)$$

which should be used to define the phase constant $\beta$, so as to meet the surface condition $\Sigma = \Delta(d\psi/dZ)_0^\pm$, giving the surface charge density $\Sigma$ in terms of the amount of discontinuity $\Delta E_0^\pm$ of the electric field strength across the surface. We get

$$\Sigma = (k_B T/e)[(d\psi/dZ)_{0+\varepsilon} - (d\psi/dZ)_{0-\varepsilon}] =$$

$$2[1/\coth(\varepsilon+\beta)][-1/sh^2(\varepsilon+\beta)] - 2[1/\coth(-\varepsilon+\beta)][-1/sh^2(-\varepsilon+\beta)] \approx$$

$$2(-\varepsilon+\beta)[1/(-\varepsilon+\beta)^2] - 2(\varepsilon+\beta)[1/(\varepsilon+\beta)^2] \approx 4\varepsilon/\beta^2 \qquad (15)$$

and setting $\varepsilon \sim \beta$ we arrive at $\beta \sim 4/\Sigma$. The reciprocal relationship between phase and surface electrostatic density agrees with the divergent character of the surface potential which requires elevating the initial point on the $\psi \to \zeta$ curve so as to match the increased $\Sigma$ and vice versa.

## 4. Conclusion

The present study was aimed at accounting for the space charges while considering the photo-charging problem, as described in our earlier arXiv papers [1-3]. Actually, the space charge was mentioned qualitatively within the context of the general (photo)electrification process but was not supported by any specific calculations leaving the matter for a further study. Now we can clearly distinguish between two extreme cases for the balance of surface and bulk charges, those of fast and slow surface-to-bulk photoequilibria. The fast ones keep the bulk always intact for any changes of the surface potential brought about by photo charging, the slow ones lead to the gradual accumulation or depletion of surface charge which is not balanced by bulk changes within the duration of an experiment. In any event, though, the bulk charges and the outer atmosphere alike play a major role in the mechanism of photo charging through controlling the surface conditions.

As shown earlier [1], the intrinsic charges in thermal equilibrium have been dealt with by the Poisson-Boltzmann approach. Such was used herein too in order to calculate both the surface density and the bulk charge distributions. The photo charging problem was again dealt with by means of the Poisson equation linking electrostatic potential to space charge, while the dependence of the concentrations on the electrostatic potentials was elaborated by the current continuity equation. The diffusing charge carriers were discarded which simplified the general problem mathematically though not physically. Future studies must incorporate diffusion too.